\documentclass[journal]{IEEEtran}
\ifCLASSINFOpdf
   \usepackage[pdftex]{graphicx}
\else
\fi
\usepackage{caption}
\usepackage{subcaption}
\usepackage[dvipsnames]{xcolor}
\usepackage{ulem}
\usepackage{soul}
\usepackage{amsmath,esint}
\begin{document}
\title{

Intermittent current interruption method for commercial lithium ion batteries aging characterization}

%\author{xx% <-this % stops a space

%\thanks{xxxx}
%}
%~\IEEEmembership{Senior Member, IEEE}
\author{Zeyang Geng, Torbj{\"o}rn Thiringer \IEEEmembership{Senior Member, IEEE}, and Matthew J. Lacey% <-this % stops a space

\thanks{Z. Geng and T. Thiringer are with the Department of Electrical Engineering, Division of Electric Power Engineering, Chalmers University of Technology, 412 96 Gothenburg, Sweden (e-mail: zeyang.geng@chalmers.se, torbjorn.thiringer@chalmers.se).

Matthew J. Lacey is with Scania CV AB, 15187 Södertälje, Sweden (e-mail:matthew.lacey@scania.com)

\textbf{This paper has been accepted for publication by IEEE. DOI (identifier) 10.1109/TTE.2021.3125418. © Copyright 2021 IEEE. Personal use of this material is permitted. Permission from IEEE must be obtained for all other uses, in any current or future media, including reprinting/republishing this material for advertising or promotional purposes, creating new collective works, for resale or redistribution to servers or lists, or reuse of any copyrighted component of this work in other works.}}

}

\maketitle

\begin{abstract}

In this article, a pioneering study is presented where the intermittent current interruption method is used to characterize the aging behaviour of commercial lithium ion batteries. With a very resource-efficient implementation, this method can track the battery resistive and diffusive behaviours over the entire state of charge range and be able to determine the aging throughout the life time of the batteries. In addition, the incremental capacity analysis can be carried out with the same data set. This method can provide measurement results with a high repeatability and produce equivalent information as the electrochemical impedance spectroscopy method. In this study, both the resistive and diffusive parameters increase with the battery capacity fading. This method does not require advanced test equipment and even with a 0.1 Hz sampling frequency, it is possible to extract usable parameters by prolonging the interruption length. Therefore, it has the potential to be easily implemented in the charging sequence in electric vehicles or stationary storage batteries for aging diagnostics.

\end{abstract}


% \fi
\IEEEpeerreviewmaketitle

\section{Introduction}

To fulfill the battery warranty to the customers, the manufacturers of electric vehicles need to predict the lifetime of their lithium ion batteries depending on the assumed usage. Despite the high cost, a massive accelerated aging test performed in a laboratory environment is still the required approach to estimate the battery lifetime \cite{ groot2012state, wikner2017lithium,wang2014degradation}. During the lifetime testing, a reference performance test (RPT) needs to be performed regularly to track the battery performance \cite{dubarry2020perspective}. An RPT often consists a series of characterization tests to obtain the important performance properties, such as capacity and resistance \cite{barai2019comparison,mulder2011enhanced}. Although there are international standards that suggest test specifications for commercial batteries \cite{international2011secondary, iso201112405, iso201212405}, there is no standard RPT procedure and instead it is designed and adapted depending on the purpose and available facility.

The most common test methods used in RPTs are the capacity test and pulse test \cite{han2014comparative,kim2019data,soto2019analysis}. The capacity test includes constant current constant voltage (CCCV) capacity test and constant current (CC) discharge capacity test. The CCCV capacity test quantifies the thermodynamic capacity of the cell and the CC discharge capacity test evaluate the usable capacity under a certain C-rate. The pulse test can be designed as different variants from the standard hybrid pulse power characterization (HPPC) test \cite{hu2012comparative}. The result from a base pulse test is an overall internal impedance,  contributed from the complex electrochemical processes, at a certain C-rate and time length \cite{ecker2014calendar}. Both the capacity test and pulse test are simple and easy to implement in an RPT procedure, but can only provide basic indicators for the battery SOH. There are other advanced characterization methods that are possible to be employed in an RPT to track more detailed aging properties, including the galvanostatic intermittent titration technique (GITT) and electrochemical impedance spectroscopy (EIS). With a sufficient long relaxation period after each current pulse, GITT can measure the open circuit voltage (OCV) of the cell including the hysteresis \cite{srinivasan2006existence}. However, to achieve an accurate result, the relaxation period should be at least one hour \cite{li2016fast} and can be up to four hours \cite{barai2015study}. The long test period introduces extra calendar aging during the lifetime test making GITT less popular in RPTs. Besides the OCV, this technique can also be used to measure the diffusion coefficient of electrode materials \cite{wen1979thermodynamic,zhu2017electrochemical,dees2009analysis}.  Compared with GITT, EIS can not only provide information about the diffusion process, which is pronounced in the low frequency range, but also the kinetic properties and ohmic resistance \cite{waag2013experimental}. Despite the challenges in interpreting the results, EIS is a very powerful tool for aging characterization \cite{zhang2011cycling,ecker2012development}. However, performing an EIS requires dedicated hardware and is thus beyond the capability of most of the testers used in a massive lifetime test. Therefore, either only a limited number of cells can be characterized with the EIS technique during aging, or the cells under test need to be manually reconnected to a different instrument to perform EIS. The re-connections will disturb the test, causing experiment errors \cite{taylor2019insight} and requiring extra time and effort. 

%The 1C discharge capacity is often used as the indicator of the battery state of health (SOH) and the end of life for a vehicle battery is defined when the 1C discharge capacity reaches 80 \% of the initial capacity \cite{groot2012state}. This impedance value can be used to evaluate the power capability of the cell and it tends to increase with the battery aging.

%In a commercial cell, the diffusion phenomenon in the two electrodes are mixed and therefore GITT can only describe the diffusion process rather than giving the actual parameter value.

Among the available test methods used in RPTs for commercial batteries, the capacity test and pulse test are practical but only provide basic aging information. GITT takes a long test time which can cause extra aging, and EIS requires a special instrument. Therefore, it is of great interest to introduce an effective characterization method which can track the battery aging behaviors in detail and can be performed with regular battery testers or even with ordinary battery charging systems. In \cite{lacey2017influence}, a new intermittent current interruption (ICI) test was used to measure the resistance values for lithium sulfur batteries. This concept has also been employed for lithium ion \cite{aktekin2018understanding,bergfelt2018varepsilon,bergfelt2020mechanically} and sodium ion \cite{mogensenattempt} battery materials. Later in \cite{chien2019simultaneous}, the ICI method was extended to capture the diffusion process in lithium sulfur batteries besides the resistance values. However, in available literature, this method has only been applied on lab scaled cells, and the validation of this method for commercial cells, has not yet been performed. Moreover, the method presented in this article, in addition, has the potential to provide more information, such as incremental capacity analysis (ICA), as well as differential voltage analysis (DVA), which has not been shown previously. Furthermore, there is a need to investigate 
its repeatability, reliability, robustness to noise and hardware requirements, to examine the potential of this method in broader application scenarios.

The specific contribution with this article is that it closes this research gap, i.e. it demonstrates how the ICI method can be used in an RPT for commercial batteries aging characterization, in a very resource effective way. This is achieved by having the following goals: 1) The demonstration of the effectiveness of the ICI method in determining and tracking a series of properties during the course of the battery aging, including the resistive and diffusion related behaviours, as well as the incremental capacity analysis. 2) The validation that the information obtained from the ICI method and the EIS test provide fully comparable results through the battery lifetime. 3) The quantification of the extra aging effect that is caused by the RPT by having a cell running RPTs continuously. 4) The proof that a high-performance test equipment is not needed, and thus the ICI method can be implemented with basic battery testers or on-board equipment in electrified vehicles or stationary storage applications, for battery system aging diagnostics.

%In this article, we demonstrate how the ICI method can be used in an RPT for commercial batteries aging characterization. With this method, the resistive and diffusion behaviour of the battery can be easily tracked during aging and an incremental capacity analysis can be performed with the same set of data. The test does not require a high performance test equipment and therefore it can be implemented with basic battery testers. To validate the information obtained from the ICI method, an EIS and a pulse test are included in the same RPT and the results are compared. Moreover, the extra aging effect caused by the RPT is evaluated by having a cell running RPTs continuously through the lifetime.  

\section{Method description}

%\begin{figure}
%    \centering
%    \includegraphics{figures/Cycling_procedure.pdf}
%    \caption{Cycling procedure.}
%    \label{fig:Cycling_procedure}
%\end{figure}

The intermittent current interruption (ICI) method was proposed to monitor the battery status for photovoltaic systems \cite{kim1997monitoring} and then applied for lithium sulfur battery characterization \cite{lacey2017influence}. In this method, a low current rate is applied to charge and discharge the battery and the current is interrupted with short pulses after long intervals. In this way it can be said that the current pattern in an ICI method is reversed from the current in a GITT method. One example of an ICI test is shown in the figure below where the current is interrupted for 10 s every 5 mins during a C/5 charging event.

\begin{figure}[!ht]
    \centering
    \includegraphics{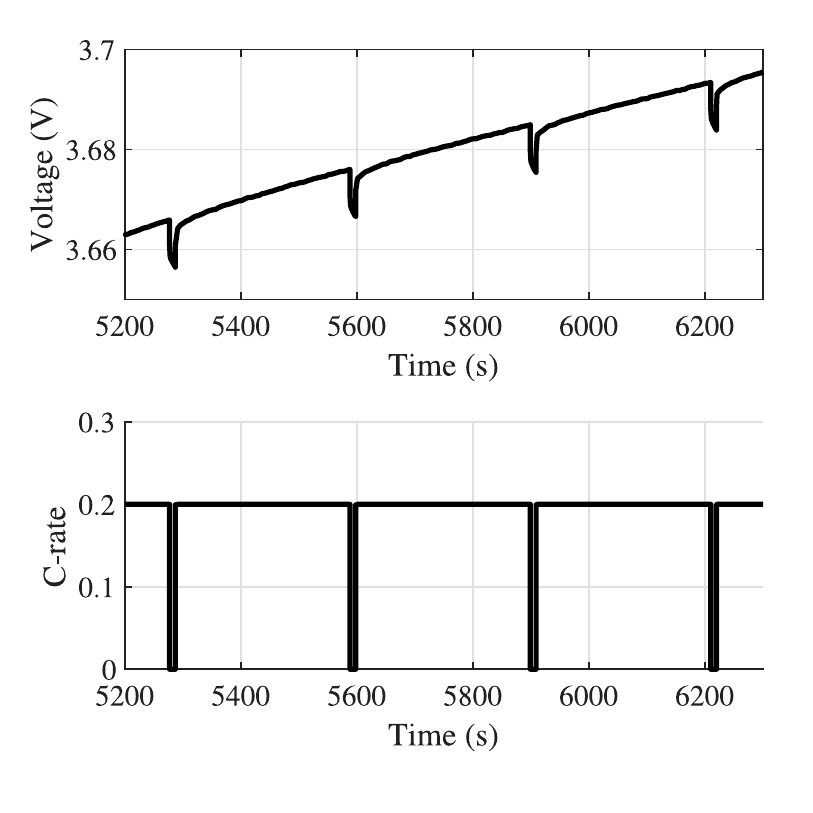}
    \caption{An example of the ICI method implementation, where a C/5 charging current is interrupted for 10 s every 5 mins.}
    \label{fig:ICI_procedure}
\end{figure}

\begin{figure}[!ht]
    \centering
    \includegraphics{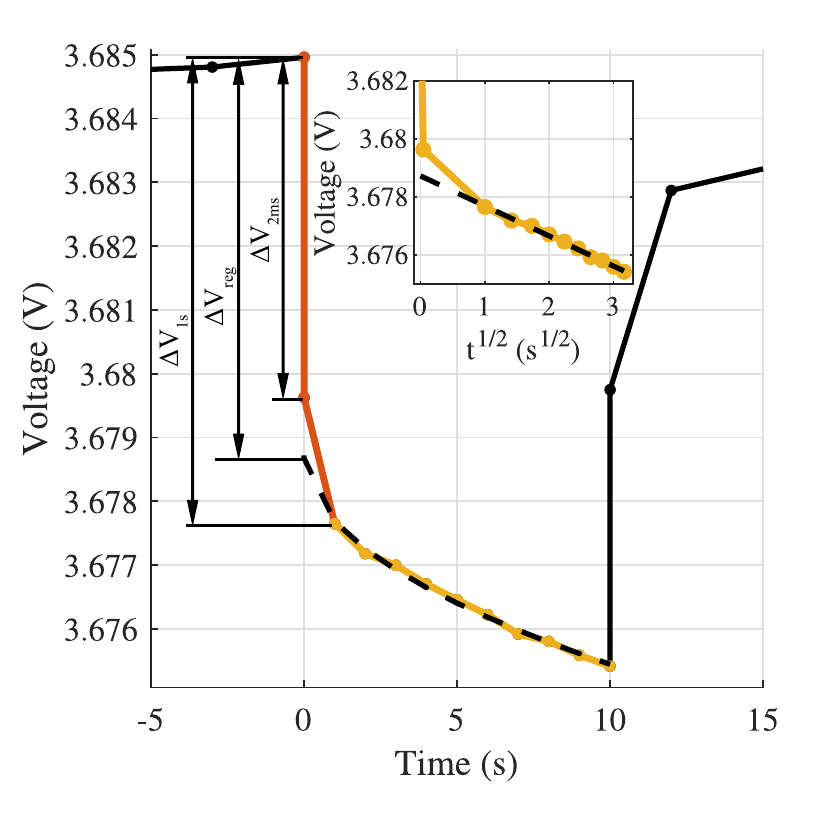}
    \caption{The voltage response at different time constants after the current is interrupted during charging.}
    \label{fig:ICI_one_pulse}
\end{figure}

After the current is interrupted during charging, the voltage drops quickly during the first second, and then decreases with a linear relationship with the square root of the time, as shown in Fig.~\ref{fig:ICI_one_pulse}. The voltage responses with different time constants can be related with different electrochemical processes taking place in a battery system. A commercial battery system has an inductive behaviour at a very high frequency (kHz) which is rarely captured in a pulse measurement at $t=$ 0 s. At $t =$ 2 ms, an instantaneous voltage drop $\Delta V_{2ms}$ can be observed which is related with the ionic resistance of the bulk electrolyte (due to the migration and not the diffusion process), as well as the electronic resistance of the electrodes and the current collectors. With time passing, the charge transfer reactions appear in the voltage response $\Delta V_{1s}$ at $t =$ 1 s. Based on the voltage change, the resistive parameters $R_{2ms}$ and $R_{1s}$ can be calculated as
\begin{equation}
    R_{2ms} = -\frac{\Delta V_{2ms}}{I},
\end{equation}
\begin{equation}
    R_{1s} = -\frac{\Delta V_{1s}}{I},
\end{equation}
where $I$ is the current before the interruption. 
For a battery tester that has a moderate sampling frequency, it is more practical to capture the total voltage change $\Delta V_{1s}$ at $t=$ 1 s.

The charge transfer reaction can be described with the Butler–Volmer equation
\begin{equation}
       j = j_0 (exp\frac{\alpha_a F \eta}{R T} - exp \frac{-\alpha_c F \eta}{R T}) 
    \label{eq:B-V_equation}
\end{equation}
and the charge transfer resistance at the particle surface $R^{'}_{ct}$ is
\begin{equation}
       R^{'}_{ct} = \frac{\eta}{j}
        \label{eq:Rct}
\end{equation}
where $j$ is the charge transfer current density per surface area, $j_0$ is the exchange current density, $\alpha_a$ and $\alpha_c$ are the anodic and cathodic constant, $F$ is the Faraday constant, $R$ is the gas constant, $T$ is temperature and $\eta$ is the overpotential caused by the redox reaction. In \eqref{eq:Rct}, the unit of $R^{'}_{ct}$ is [$\Omega \cdot$ m$^2$] and the area here is the total surface area of the particles. In commercial lithium ion batteries, the electrode material has a porous structure to increase the contact surface area between the particle and the electrolyte. The charge transfer resistance of the electrode $R_{ct}$ is
\begin{equation}
    R_{ct} = \frac{R_{ct}^{'}}{A d S_a }
\end{equation}
where $A$ is the electrode area, $d$ is the electrode thickness and $S_a$ is the specific surface area, which is the total surface area per volume. In a physical model of the lithium ion battery,  $R_{ct}$ is connected with the double layer capacitor in parallel, and this RC link dynamic behaviour dominates the voltage response between 0 s and 1 s.

After 1 s, the voltage response is governed by the diffusion process in the solid as it is the slowest process, and this is similar in a GITT test. The diffusion process is the solid can be described with Fick's laws of diffusion in a spherical coordinate system
\begin{equation}
\frac{\partial c_s}{\partial t} = D_s (\frac{\partial ^2 c_s}{\partial r^2} + \frac{2}{r} \frac{\partial c_s}{\partial r})
       \label{eq:Ficks_law}
\end{equation}
where $t$ is time, $r$ is the distance from the particle center, $c_s$ is the lithium ion concentration in the solid and $D_s$ is the diffusion coefficient. The boundary condition is
\begin{equation}
\frac{\partial c_s}{\partial r}|_{r = r_s} = -j/(FD_s)
\end{equation}  
where $r_s$ is the particle radius, $j$ is the current density at the particle surface and $F$ is the Faraday constant. The symmetry of the particle also gives
\begin{equation}
\frac{\partial c_s}{\partial r}|_{r=0} = 0
\end{equation}
When the current is interrupted, the voltage response $V$ follows a linear relationship of $\sqrt{t}$
\begin{equation}
     V = V_0 + \frac{dE}{dc} \frac{j}{F} \sqrt{\frac{t}{D_s}},~t \ll \frac{r_s^2}{D_s}
     \label{eq:GITT}
\end{equation}
A diffusion related parameter $k$ can be extracted
\begin{equation}
   k = -\frac{1}{I} \frac{dV}{dt^{1/2}},~t>1s.
\end{equation}
In \eqref{eq:GITT} the diffusion coefficient $D_s$ is a parameter for a specific electrode material. However when measuring a commercial battery cell, it is not possible to distinguish the impact of the positive electrode and the negative electrode. Therefore the parameter $k$ reflects the diffusion behaviour of the two electrodes.

The $k$ value obtained from the ICI method can be correlated with the diffusion parameter $\sigma$ extracted from the Warburg impedance $Z_w = \sigma \omega ^{-1/2} - j\sigma \omega ^{-1/2}$ in an EIS measurement, where $\omega = 2 \pi f$ is the angular frequency. The relationship between $\sigma$  and $k$ has been theoretically derived in \cite{chien2019simultaneous}, where it was found to be
\begin{equation}
   k = \sigma \sqrt{\frac{8}{\pi}}.
\end{equation}

The time constants for different electrochemical processes depend on the materials in the battery. The 2 ms and 1 s above were selected arbitrarily based on the priori knowledge from the impedance spectrum of the battery used in this work. To remove the arbitrariness of the timescale, a linear regression analysis can be carried out, which was proposed in \cite{lacey2017influence}. If the linear regression of the voltage response is backwards extrapolated to $t=$ 0, a voltage change $\Delta V_{reg}$ can be calculated as shown in Fig.~\ref{fig:ICI_one_pulse}. The corresponding resistance is 
\begin{equation}
    R_{reg} = -\frac{\Delta V_{reg}}{I}.
\end{equation}
This value $R_{reg}$ can be accurately obtained even when the test is performed in a noisy environment or with a poor sampling rate. Therefore this parameter can be used to track the battery aging characters in field diagnostic tests. Moreover, by tracking $R_{reg}$ instead of the resistance at a specific time, the impact of various timescales can be eliminated.

Beside the resistive and diffusive parameters, the ICA and DVA can be performed at the same time. The outcome of the ICA is a useful description of the battery aging status \cite{dubarry2011evaluation}.

The results from the ICI method can be affected by the current magnitude, With a lower current, the battery is closer to the equilibrium state during the test, providing more detailed ICA and DVA curves, but on the other hand, this unfortunately leads to a longer test time. With a higher current, the charge transfer resistance in \eqref{eq:Rct} decreases and the diffusion parameter $k$ is also affected by the concentration polarization built up in the electrolyte. The C/5 current rate selected in this work is a trade off between testing time and validity of the extracted information.  
%\begin{figure}
%    \centering
%    \includegraphics{figures/EIS_OCV.pdf}
%    \caption{EIS OCV}
%    \label{fig:EIS_OCV}
%\end{figure}

\section{Experiment results and analysis}

In this section, the method's repeatability is verified first and then the ICI method is included in the RPT sequence during a lifetime test to demonstrate how the ICI method can be used to track the battery aging phenomenon. During the lifetime, an EIS is performed after each ICI to show the correlation between the two methods. In parallel, a cell has been cycled with the RPT sequence only to evaluate the aging effect introduced by the RPT.

The test objects used in this work are 26 Ah lithium ion pouch cells with a voltage window of 2.8 V - 4.15 V. The positive electrode is a mixture of LiNi$_{0.33}$Mn$_{0.33}$Co$_{0.33}$O$_2$ and LiMn$_2$O$_4$ and the negative electrode is graphite. During the test, the cells are placed in a test jig with a certain pressure applied. All the tests are performed in a temperature chamber set at 25 $^o$C. The cycling equipment is a PEC ACT0550.

\subsection{Repeatability of the ICI method}

\begin{figure}
    \centering
    \includegraphics{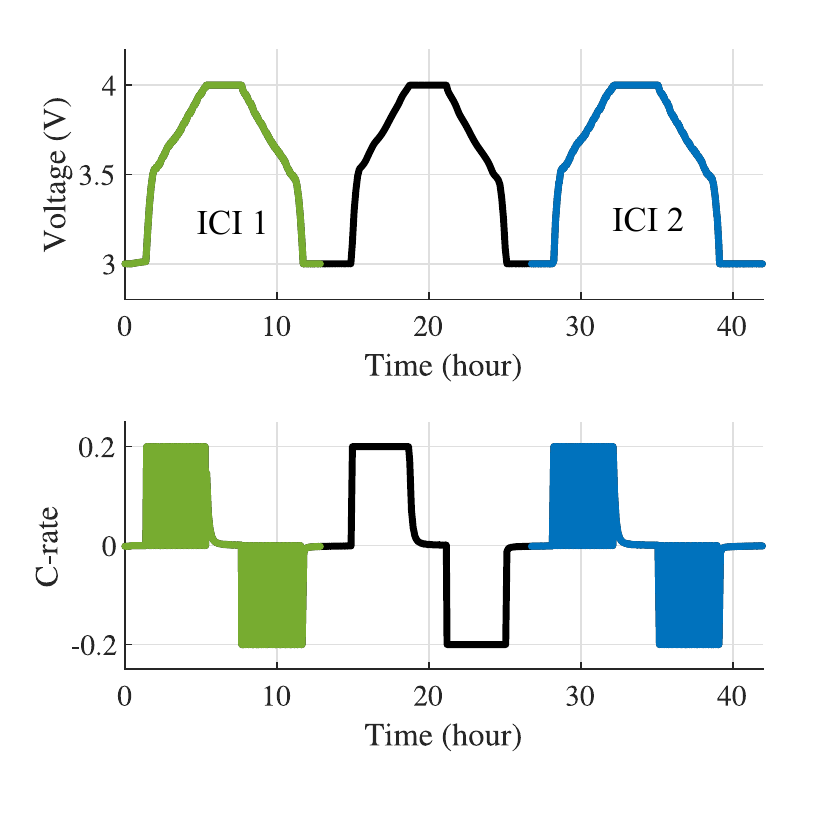}
    \caption{Two ICI tests are performed with a regular CCCV cycle in between to examine the repeatability of the method. Zoom-in details are shown in Fig.~\ref{fig:ICI_procedure} and Fig.~\ref{fig:ICI_one_pulse}. }
    \label{fig:ICI_repeatability}
\end{figure}

\begin{figure}
    \centering
    \includegraphics[scale = 0.9]{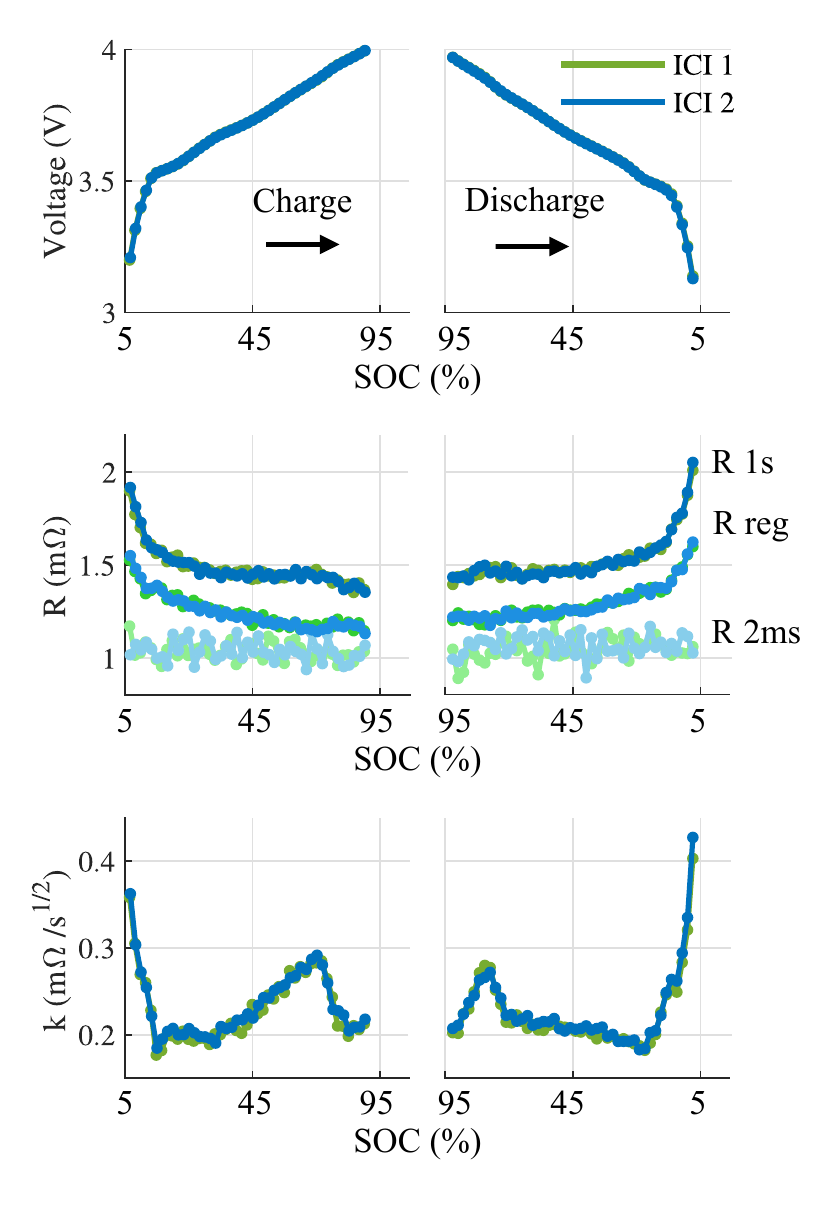}
    \caption{The parameters $R_{2ms}$, $R_{reg}$, $R_{1s}$ and $k$ obtained from the two ICI tests in Fig.~\ref{fig:ICI_repeatability} showing a high repeatability.}
    \label{fig:ICI_repeatability_result_r_k}
\end{figure}

\begin{figure}
    \centering
    \includegraphics{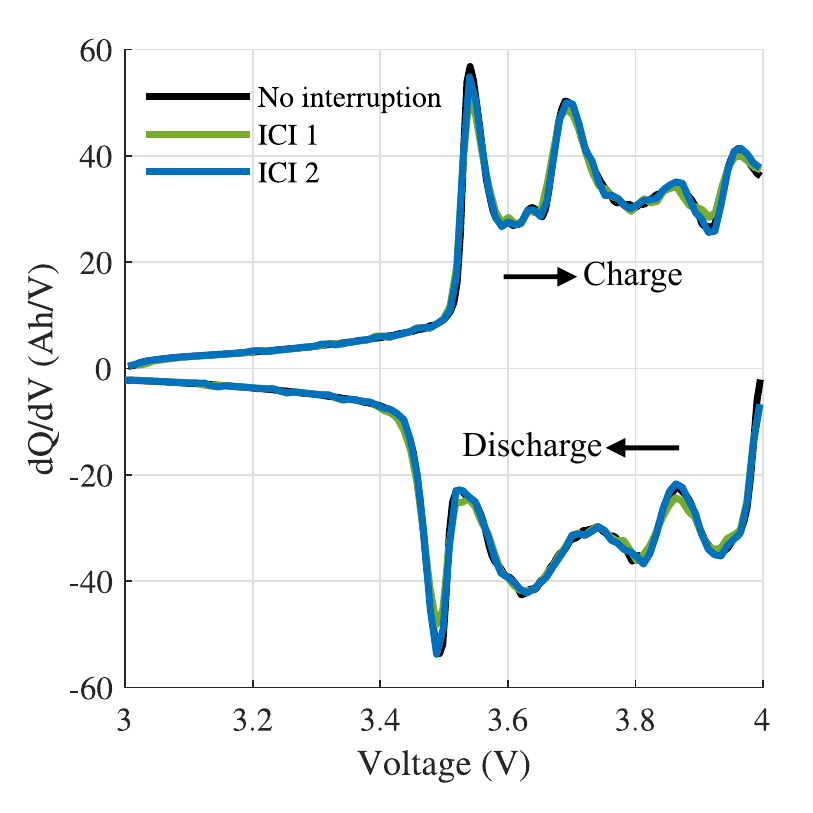}
    \caption{The incremental capacity analysis based on the two ICI tests in Fig.~\ref{fig:ICI_repeatability} and a regular CCCV cycle without current interruption.}
    \label{fig:ICI_repeatability_result_ICA}
\end{figure}

In Fig.~\ref{fig:ICI_repeatability}, a test procedure is implemented to verify the repeatability of the measurement result obtained from the ICI method, where a regular CCCV cycle is applied in between two ICI tests. The zoom-in details are shown in Fig.~\ref{fig:ICI_procedure} and Fig.~\ref{fig:ICI_one_pulse}. The results obtained from the two ICI tests are presented in Fig.~\ref{fig:ICI_repeatability_result_r_k} and Fig.~\ref{fig:ICI_repeatability_result_ICA}. The excellent repeatability of the experimental result shows that the method can be used to characterize the battery aging with a high reliability.

It can be observed in Fig.~\ref{fig:ICI_repeatability_result_r_k} that the pure resistive parameter $R_{2ms}$ is independent from the state of charge (SOC) since the ionic resistance of the bulk electrolyte and the electronic resistance of the electrodes and current collector are not affected by the SOC. On the other hand, $R_{reg}$ and $R_{1s}$ show a clear increasing trend towards the low SOC range. The diffusion related parameter $k$ is higher at the low SOC, indicating a decreased diffusion coefficient. Moreover, the $k$ value has a distinct peak around 70 \% SOC. In both the resistive parameter $R_{reg}$ and $R_{1s}$ and the diffusive parameter $k$, a slight difference can be observed in the parameter values obtained during the charging process and discharging process.

Fig.~\ref{fig:ICI_repeatability_result_ICA} shows the ICA plot of the battery. The peaks in the ICA plot correspond to the plateaus in the voltage profile, which are signs of the phase changes in the electrode materials. As can be noted from the figure, the ICA plot obtained from the ICI data agrees very well with the plot obtained from a regular CCCV cycle. This shows that the current interruptions in the ICI method does not introduce significant artefacts in the analysis.

\subsection{Tests at different temperatures}

\begin{figure}
    \centering
    \includegraphics{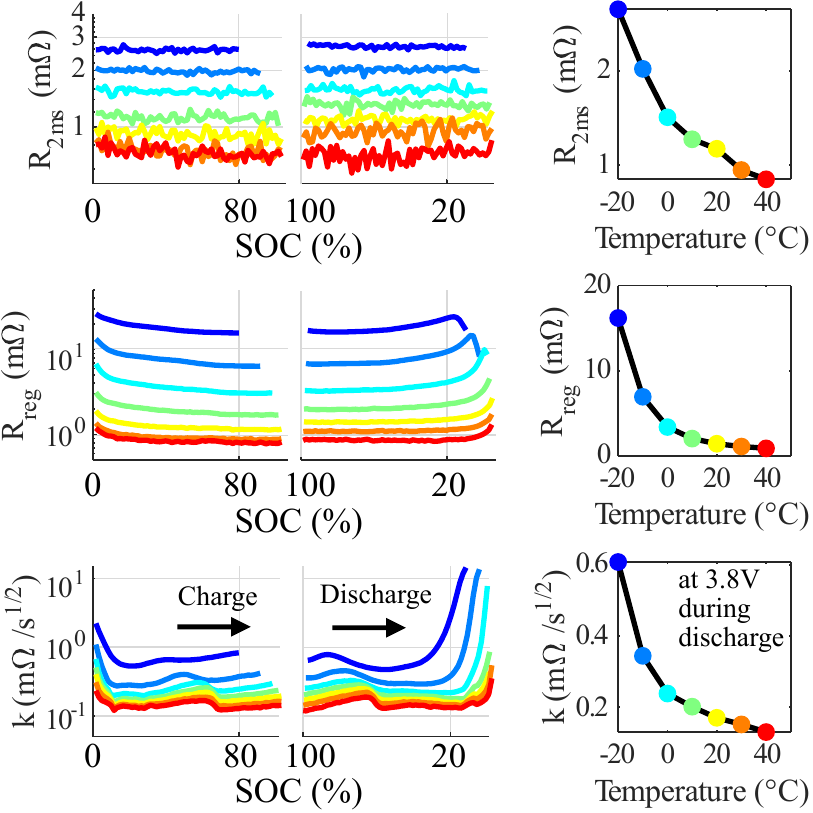}
    \caption{Results of ICI tests performed at different temperatures from -20 $^o$C to 40 $^o$C.} 
    \label{fig:results_diff_temps}
\end{figure}

Lithium-ion batteries can operate over a wide temperature range and its electrical properties are highly dependent on the temperature. In this section, the validity of the ICI method at different temperatures are investigated. The temperature condition of the test is from -20$^o$C to 40 $^o$C, with a step of 10 $^o$C, and it covers most of the temperature operation conditions for batteries in electric vehicles. The battery cell under test is located in an ESPEC LU-124 climate chamber. Before each test, the battery is left in the desired ambient temperature for at least 8 hours to allow it to reach a thermal equilibrium state and the cell is CCCV discharged to its minimum SOC. The test procedure is a 10 s pulse every 5 mins during a C/5 charge and discharge, the same in all temperature conditions. The test results are shown in Fig.~\ref{fig:results_diff_temps}. The figures on the left present the data over the entire SOC range at different temperatures and the color coding is according to the figures on the right. The y-axis in the three figures on the left are in logarithmic scale to provide a clear visualization. From the test at each temperature, the electrical properties, $R_{2ms}$, $R_{reg}$ and $k$ measured at 3.8 V during discharge are selected and plotted versus the temperature in the figures on the right, with a linear scale on the y-axis. There is an underestimation of the $R_{reg}$ values when the SOC is lower than 10 \%, at the very low temperature conditions, i.e. -20 $^o$C and -10 $^o$C, where the voltage response is not diffusion controlled anymore during the interruptions. It can be seen that at a lower temperature, both the resistive parameters $R_{2ms}$ and $R_{reg}$ and the diffusive parameter $k$ are higher. This is because the electrolyte conductivity, exchange current density and diffusion coefficients (both in the particle and in the electrolyte) decrease at lower temperature, resulting in higher resistance values. This trend appear at the entire SOC range, both during charge and discharge. The test results show that the ICI method can be applied at different temperatures to characterize batteries without adjusting the design of the test. Despite the fact that the time constant of charge transfer reaction varies at different temperature, the linear regression approach removes the arbitrariness of the timescale, and one can obtain essentially time-independent quantities to describe the resistance of the system.

\subsection{Aging test}

\begin{figure}
    \centering
    \includegraphics{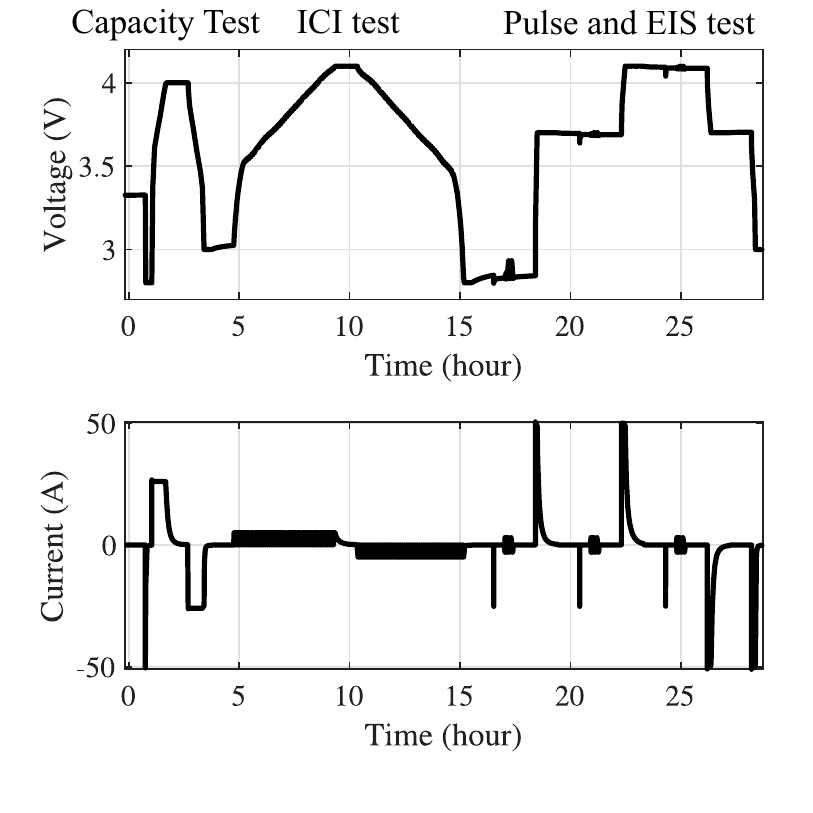}
    \caption{The RPT sequence used in the aging test, including a 1C capacity test, ICI test, 1 kHz AC impedance test, EIS test and pulse tests.}
    \label{fig:RPT_example}
\end{figure}

During the aging test, the battery cell has been cycled with +2C/-2C constant current in a voltage window of 3 V - 4 V at 25 $^o$C. An RPT is scheduled regularly to track the battery performance. The RPT sequence is shown in Fig.~\ref{fig:RPT_example}, including a 1C capacity test, ICI test, 1 kHz AC impedance test, EIS test and pulse tests. The 1C capacity is measured between 3 V and 4 V and the ICI test is implemented with C/5 current and 5 s interruption every 5 mins. A sequence of 1 kHz AC impedance test and EIS test is performed at 2.8 V, 3.7 V and 4.15 V respectively.  2 C charge pulse tests are performed at 2.8 V and 3.7 V, and 2 C discharge pulse tests are performed at 4.15 V and 3.7 V.

\begin{figure}
    \centering
    \includegraphics{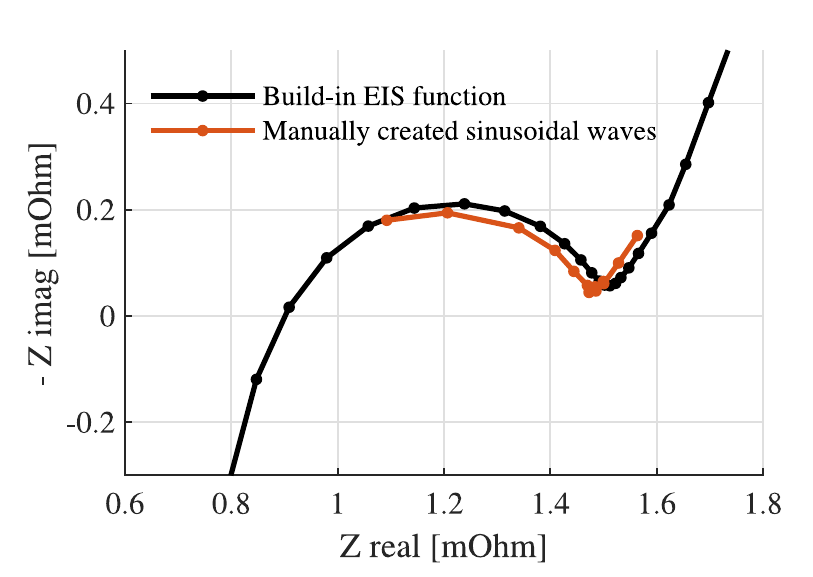}
    \caption{Verification of the EIS result obtained with a manually created sine waves pattern.}
    \label{fig:EIS_verification_GAMRY}
\end{figure}

\begin{figure}
    \centering
    \includegraphics{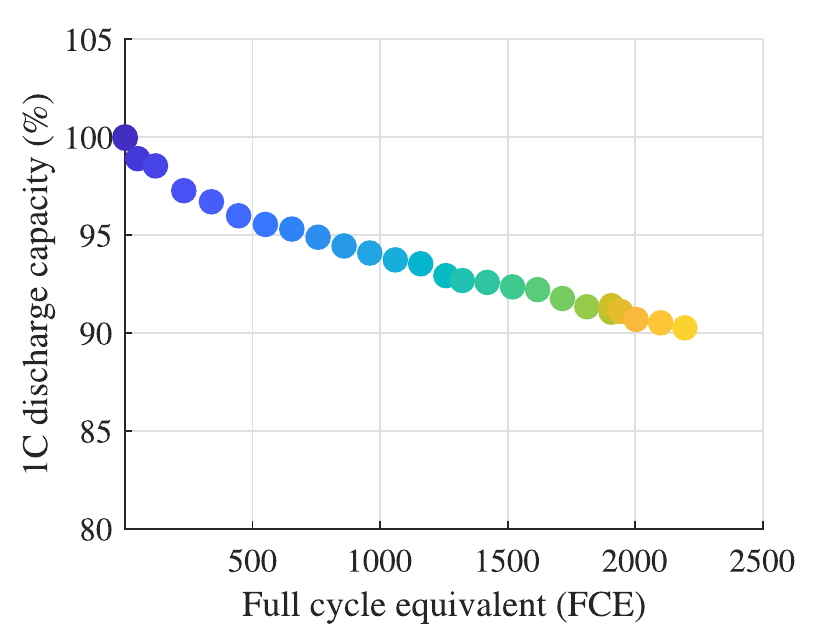}
    \caption{The degradation of the 1C discharge capacity of the investigated cell.}
    \label{fig:Aging_1C_discharge_capacity}
\end{figure}

The tester used in the aging test does not have a built-in EIS function, and therefore a current pattern consisting sinusoidal waves with different frequencies is generated in MATLAB and implemented as a drive cycle in the tester. The highest update rate and sampling rate of the input current is 1 kHz and the maximum frequency of a sine wave that can be generated without too much distortion is thus approximately 100 Hz. This approach has been verified with a potentiostat GAMRY Reference 3000 which is dedicated for EIS measurement. The impedance result measured with the GAMRY built-in EIS function and with the manually generated sine waves are compared in Fig.~\ref{fig:EIS_verification_GAMRY}. It shows that the manually created sine wave pattern can be used to extract the impedance information and it can provide a trustworthy result within a limited frequency range. This method can be applied in general to obtain an EIS plot within a limited frequency range even the battery tester that does not have a built-in EIS functionality. Another alternative is to use a pseudo-random binary sequences (PRBS) which has been proved in \cite{geng2018board}.

\begin{figure}[!ht]
    \centering
    \includegraphics{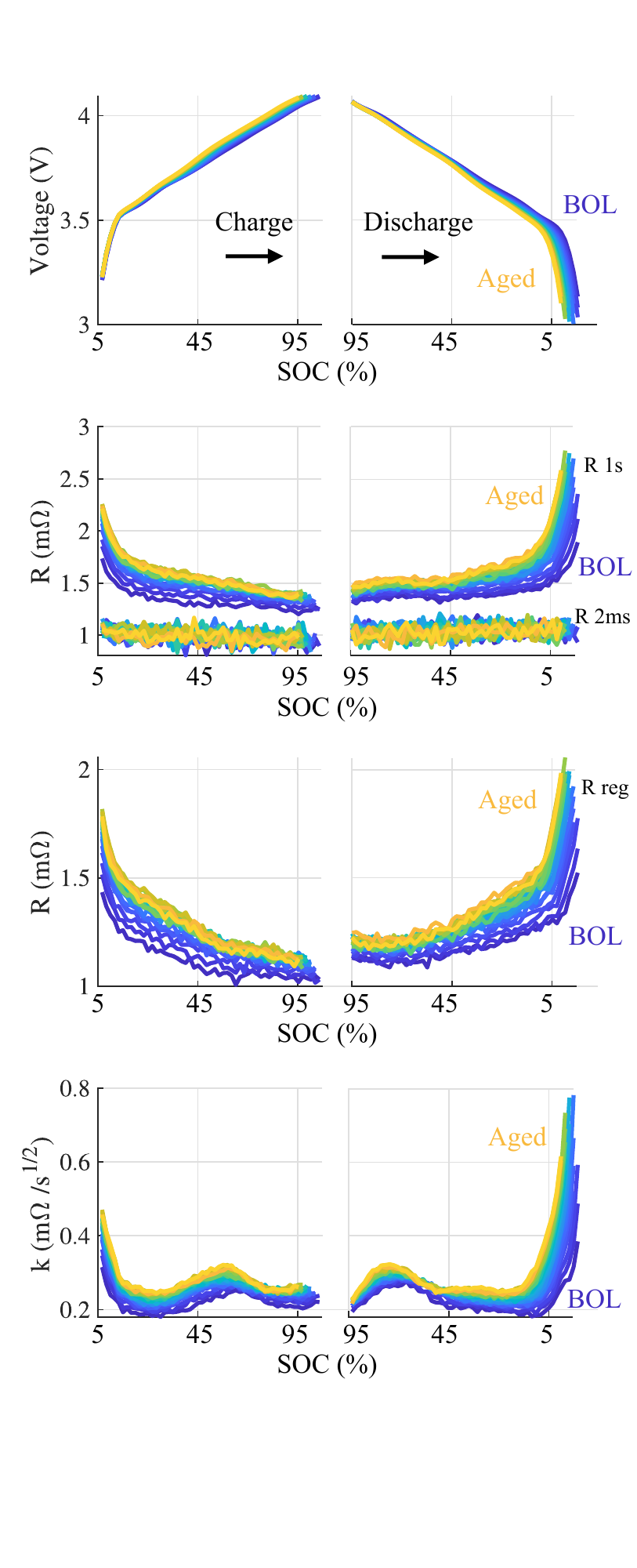}
    \caption{The resistive and diffusive parameters obtained from the ICI method during the battery aging.}
    \label{fig:Aging_ICI_resistance}
\end{figure}

The degradation of the cell 1C discharge capacity is shown in Fig.~\ref{fig:Aging_1C_discharge_capacity} and each dot represent one RPT. The 1C discharge capacity at the beginning of life (BOL) is set as 100 \% and it has only decreased to 90 \% after 2000 full cycle equivalent (FCE). The cell used in this work is a commercial cell for electric vehicle application and it has an excellent performance as well as lifetime.

\begin{figure}[!ht]
    \centering
    \includegraphics[scale = 0.9]{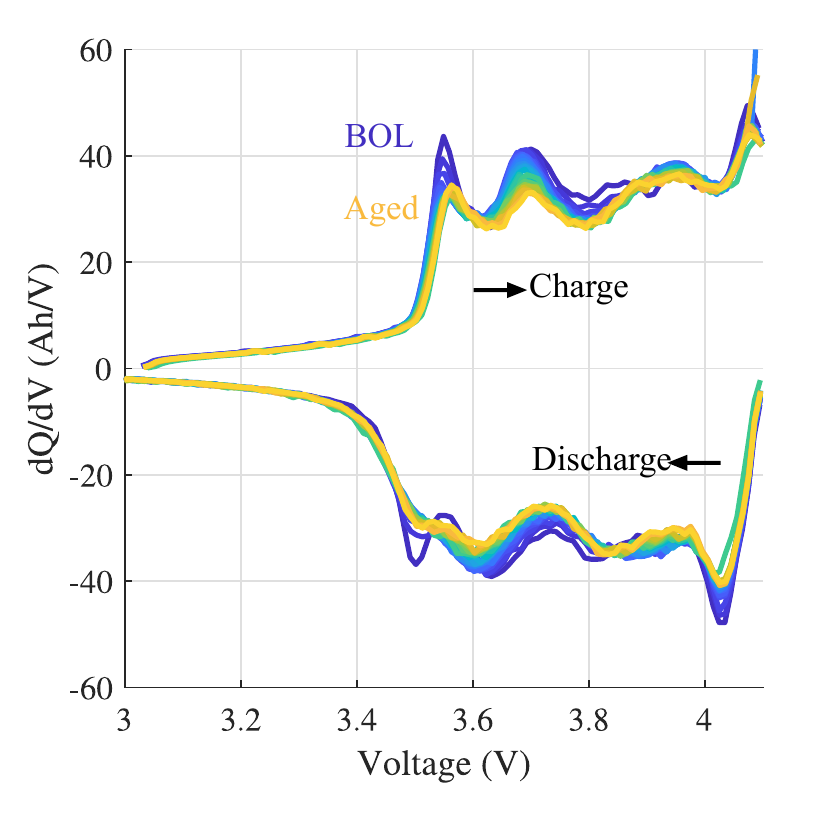}
    \caption{The incremental capacity analysis with the ICI data to track the battery degradation.}
    \label{fig:Aging_dQdV}
\end{figure}

From the ICI tests in the RPT, the extracted properties are presented in Fig.~\ref{fig:Aging_ICI_resistance} and Fig.~\ref{fig:Aging_dQdV}. The pure resistive parameter $R_{2ms} $ remains at the same level through the battery lifetime, implying that there is little degradation in the electrolyte which results in any notable change in electrolyte resistance, and no significant corrosion in the current collectors. On the other hand, there are significant increases in both $R_{reg}$ and $R_{1s}$ for the entire SOC range. These two parameters include the pure resistive component, as well as the charge transfer resistance and the solid electrolyte interphase (SEI) resistance. Since $R_{2ms}$ shows a constant behaviour, it indicates that the increases in $R_{reg}$ and $R_{1s}$ are possibly due to changes taking place at the electrodes, with a major contributor expected to be the growth of the SEI layer at the negative electrode as is typically observed for lithium-ion batteries \cite{wikner2017lithium}. Similarly, the diffusion parameter $k$ rises with the cell degrading, meaning a slower diffusion process for an aged battery. This can also be resulted by an increased SEI layer and a lower specific surface area. It can be observed that the parameter increases are more pronounced at the lower SOC range, although the very low SOC range has not been utilized during the cycling aging.

\begin{figure}
    \centering
    \includegraphics{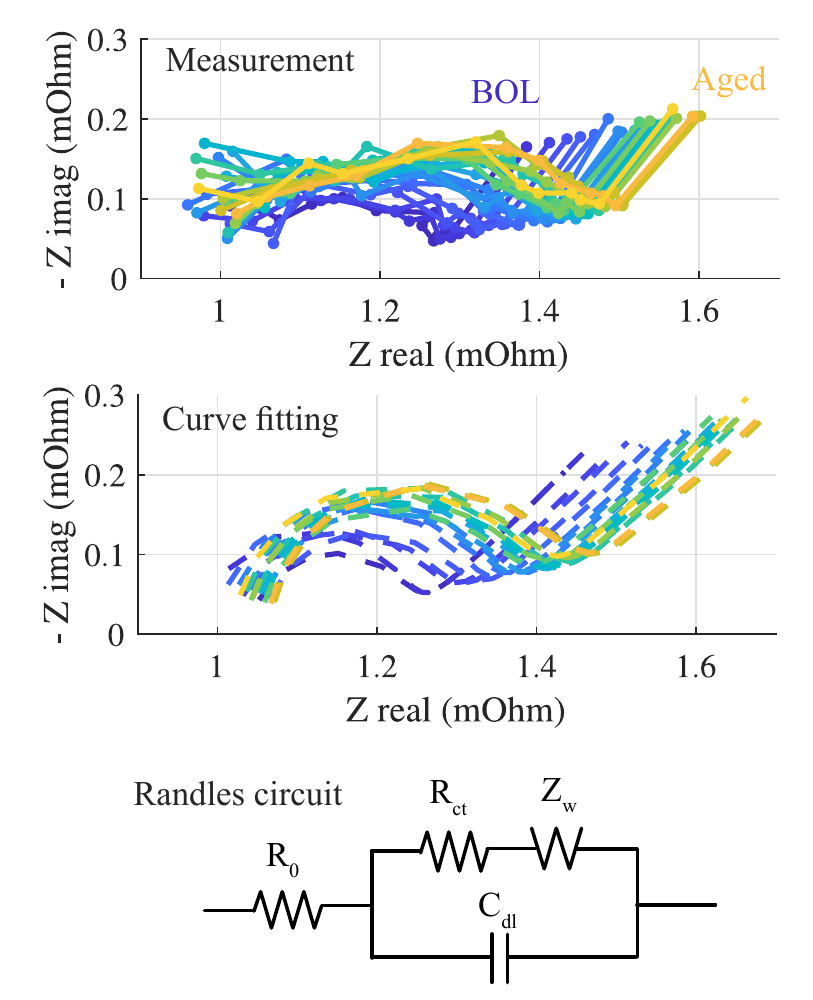}
    \caption{The cell impedance (100 Hz - 50 mHz) at 3.7 V increases with aging.}
    \label{fig:Aging_EIS_3_7V_fitting}
\end{figure}

\begin{figure}
    \centering
    \includegraphics{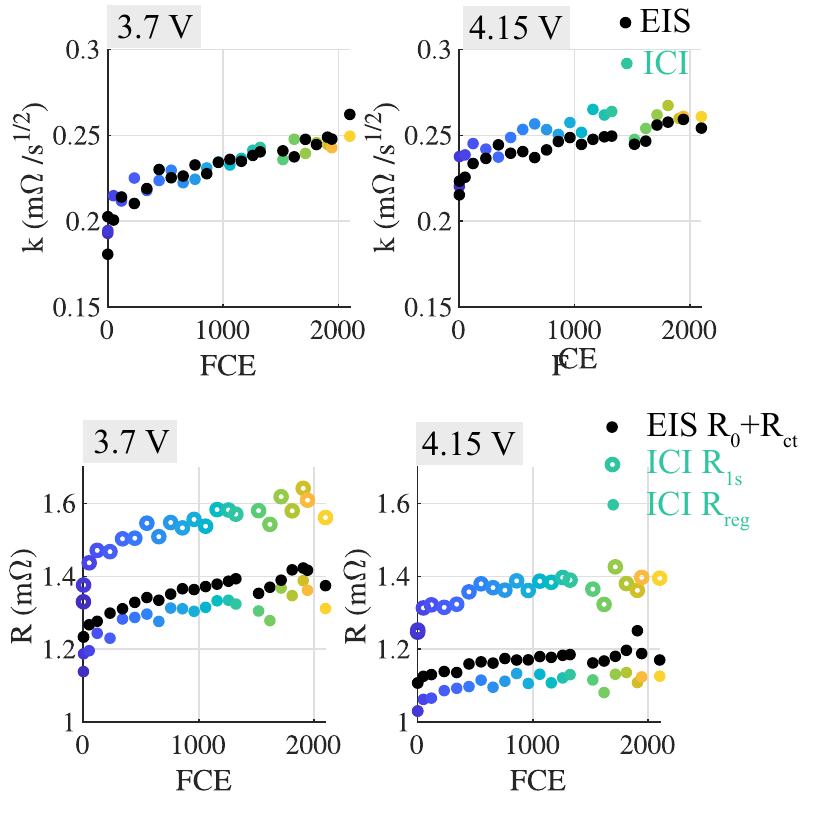}
    \caption{A comparison of the parameters obtained from the ICI method and EIS method.}
    \label{fig:Aging_EIS_ICI_fitting}
\end{figure}

As mentioned in (11), the ICI method can provide similar diagnostic information as the EIS method. In this aging test, the battery impedance between 100 Hz and 50 mHz is measured with a series of sinusoidal current waves as the input at different voltage levels.  The impedance measurements are fitted with the Randles circuit, where $R_{0}$ represents the sum of the electronic and ionic resistances, $R_{ct}$ stands for the charge transfer resistance, $C_{dl}$ is the double layer capacitance and $Z_w = \sigma \omega ^{-1/2} - j\sigma \omega ^{-1/2}$ is the Warburg impedance resulting from the diffusion processes. The diffusion parameter $\sigma$ can be extracted from the Warburg impedance and then be translated to $k$ according to (4).

In Fig.~\ref{fig:Aging_EIS_3_7V_fitting}, the impedance increase at 3.7 V resulting from the cycling aging, as well as the curve fittings, are presented. $R_{0}+R_{ct}$ and $k$ extracted from the EIS measurement at 3.7 V and 4.15 V are plotted in Fig.~\ref{fig:Aging_EIS_ICI_fitting} (black dots), showing the expected behaviours, that both parameters increase during the cycling aging. In the same plot, the parameters $R_{reg}$, $R_{1s}$, and $k$ obtained from the ICI method, are compared at two voltage levels. There is a very good agreement in the $k$ values obtained from the EIS and ICI method during the entire lifetime test at both voltage levels. Meanwhile, the EIS parameter $R_{0}+R_{ct}$ values match with the ICI parameter $R_{reg}$. Although the extracted parameters contain mixed information from the two electrodes, they are related to the actual physical processes and thus are useful indicators for the degradation phenomena. Fig.~\ref{fig:Aging_EIS_ICI_fitting} shows that the ICI method can provide equivalent information as the EIS method regarding the resistance and diffusion parameters and the ICI can provide this information over the entire SOC range as demonstrated in Fig.~\ref{fig:Aging_ICI_resistance}. 

\begin{figure}[!ht]
    \centering
    \includegraphics[scale = 0.9]{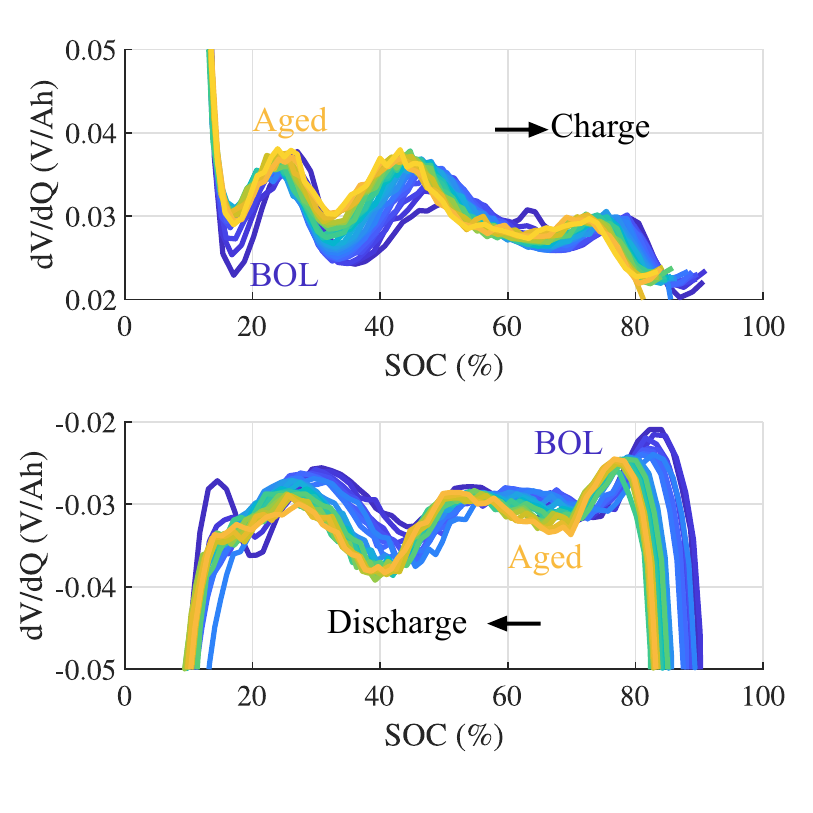}
    \caption{The differential voltage analysis with the ICI data.}
    \label{fig:Aging_DVA}
\end{figure}

Besides tracking the resistance and diffusion parameters, the ICI method can provide data for the incremental capacity plot in Fig.~\ref{fig:Aging_dQdV} and the differential voltage analysis in Fig.~\ref{fig:Aging_DVA} , which reveal where the capacity loss occurs. A decreased peak in Fig.~\ref{fig:Aging_dQdV} is indicating a reduced capacity at the corresponding phase.

%\begin{figure}
%    \centering
%    \includegraphics{figures/Aging_EIS_3_7V.pdf}
%    \caption{The cell impedance (100 Hz - 2 mHz) at 3.7 V increases with aging.}
%    \label{fig:Aging_EIS_3_7V}
%\end{figure}

%\textcolor{blue}{Within one period of the sinusoidal wave, the battery state of charge changes slightly, therefore the measured voltage change during the sweep is caused not only by the impedance but also by the open circuit voltage (OCV) change. The OCV change will affect the result more significantly at a lower frequency with a longer sweep period. In Fig.~\ref{fig:Aging_EIS_3_7V}, the OCV background change is subtracted with a defined OCV look up table (updated during aging) and the dashed lines show the impedance result after the OCV subtraction. It shows that the OCV background change makes a difference for the imaginary part of the impedance at the very low frequency range. The impact of the OCV on the impedance depends the shape of the voltage profile...?}

%\begin{figure}
%    \centering
%    \includegraphics{figures/Aging_EIS_ICI.pdf}
%    \caption{A comparison of the parameters obtained from the ICI method and EIS method. The ICI method can provide an equivalent characterization as the EIS method during a battery lifetime test.}
%    \label{fig:Aging_EIS_ICI}
%\end{figure}

\subsection{Aging effect introduced by RPTs}

\begin{figure}[!ht]
    \centering
    \includegraphics{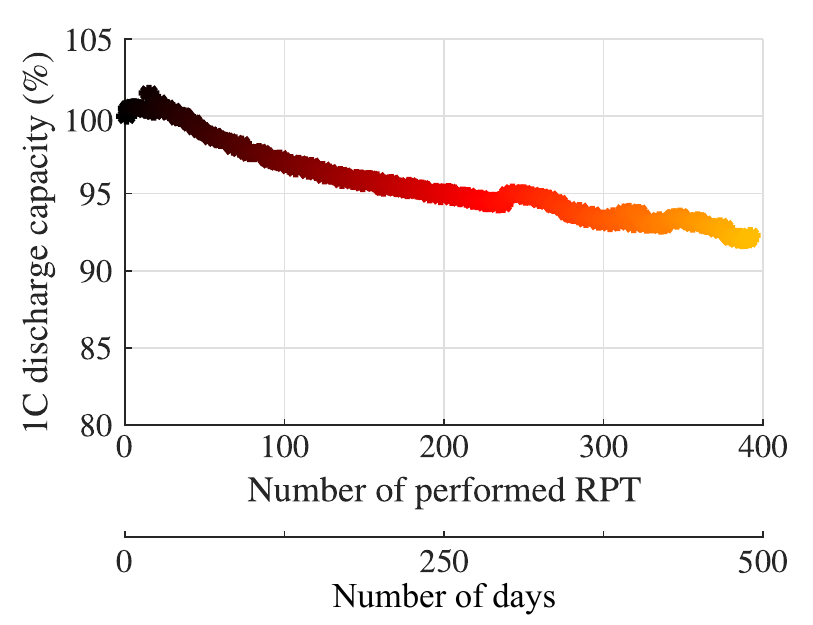}
    \caption{The cell capacity degradation with RPTs performed continuously.}
    \label{fig:Aging_continuous_RPT}
\end{figure}

It is important to evaluate and limit the extra aging introduced by the RPT sequence so that the RPT itself does not influence the cycling aging considerably. In this work, a cell of the same type used in the previous aging test has been cycled continuously with the RPT in Fig.~\ref{fig:RPT_example} and the degradation of its 1C discharge capacity is shown in Fig.~\ref{fig:Aging_continuous_RPT}. After 400 repetitive RPTs, the battery lost 8 \% of the capacity meaning that each RPT will roughly age the battery 0.02 \%. In Fig.~\ref{fig:Aging_1C_discharge_capacity}, 27 RPTs have been performed for diagnostics over 2200 FCEs, which in total can cause around 0.54 \% capacity loss. Compared with the 10 \% capacity loss during the cycling aging, the diagnostic tests introduces around 5\% extra aging, which is acceptable. This effect can be reduced further if the RPT is performed less frequent and with a simplified procedure. As shown in this article, it is possible to include only the 1 C capacity and the ICI test in the RPT, since the two tests can provide sufficient information for aging analysis.

\subsection{Analysis of the implementation requirements}

In the previous sections, it has been proved that the ICI method is able to provide high quality results with measurements obtained from an advanced battery tester (max 1 kHz sampling frequency and the voltage measurement accuracy is within $\pm$0.005 \% full scale deviation ) in a controlled lab environment. However, as indicated in the introduction of this article, the ICI method does not require high performance equipment and it can be applied with a normal tester or on-board equipment in an ordinary battery system.

\begin{figure}
    \centering
    \includegraphics{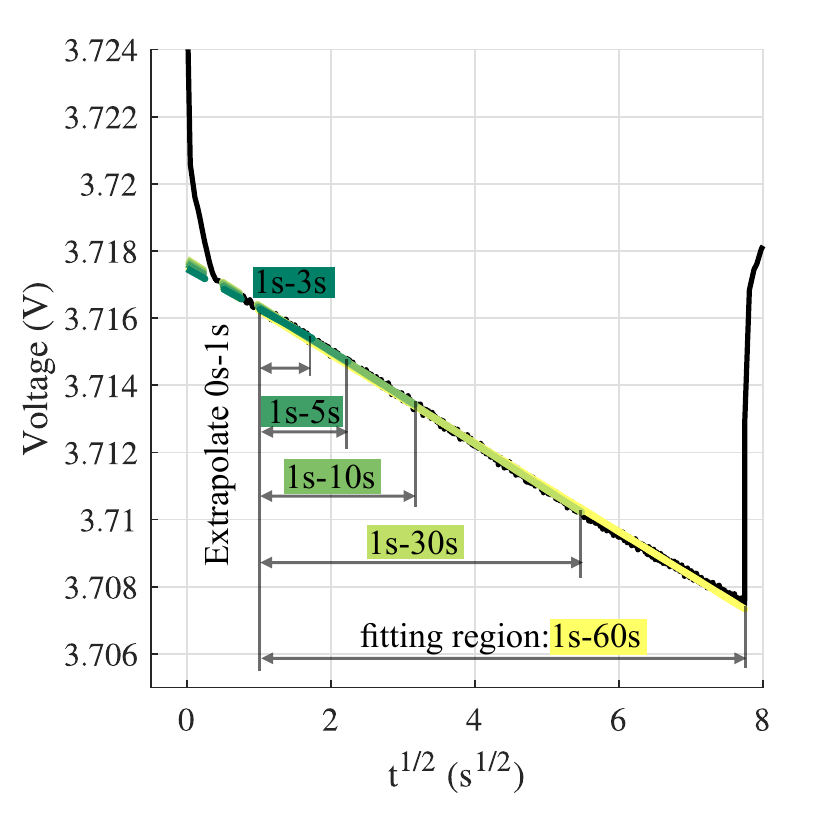}
    \caption{Different interruption lengths are used to extract the $R_{reg}$ and $k$ parameters.}
    \label{fig:pulse_different_length_fit}
\end{figure}

\begin{figure}
    \centering
    \includegraphics{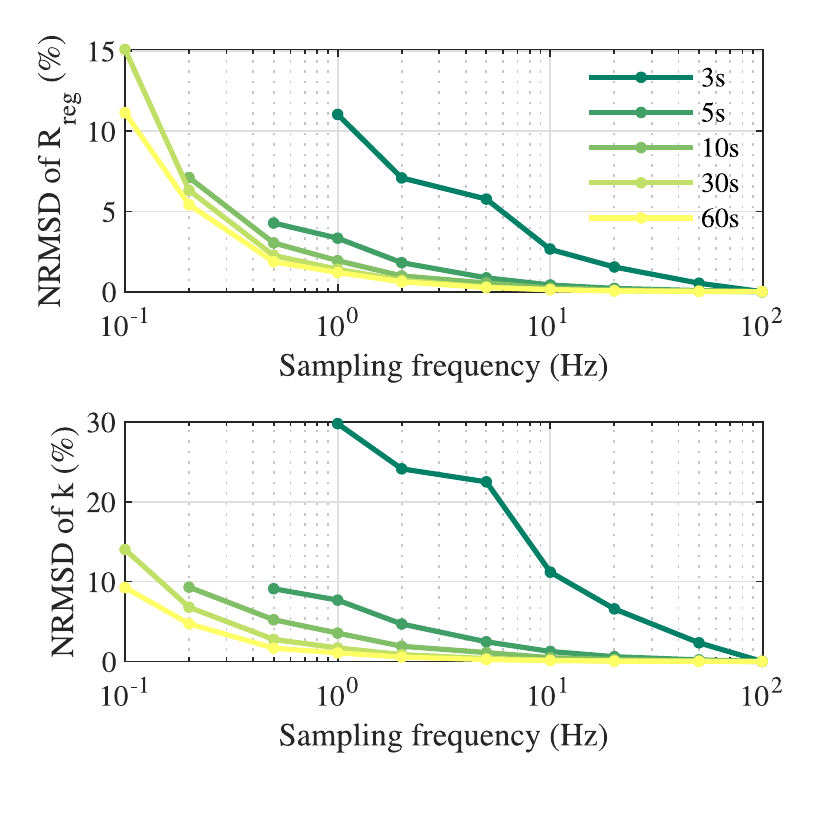}
    \caption{Normalized root mean square deviations (NRMSDs) of the results extracted with different sampling frequencies for each interruption length case.}
    \label{fig:RMSD_length_freq}
\end{figure}

In Fig.~\ref{fig:ICI_procedure} and Fig.~\ref{fig:ICI_one_pulse}, the interruption length was 10 s and the sampling frequency was 1 Hz (one extra sample at 2 ms after the current interruption) and the corresponding results in Fig.~\ref{fig:ICI_repeatability_result_r_k} and Fig.~\ref{fig:ICI_repeatability_result_ICA} showed the battery properties with clear trends. In many applications, including electric vehicles, 1 Hz or a higher sampling frequency is available, therefore the procedure described previously can be directly adapted in a charging event. However, in some other applications, for example in a stationary storage, the data acquisition system might not be able to log the data fast enough. In such cases, a longer interruption period can be adapted to extract the resistance $R_{reg}$ in (5) and a diffusion related $k$ parameter in (4). In Fig.~\ref{fig:pulse_different_length_fit}, the voltage response during a 60 s current interruption during charging is plotted versus $t^{1/2}$. For each interruption length, the result obtained from the original measurement (500 Hz sampling) is considered as the reference result. The deviations of the results obtained from a down-sampled data are shown in Fig.~\ref{fig:RMSD_length_freq} (only during charging). The normalized root mean square deviation (NRMSD) is calculated as
\begin{equation}
RMSD (f_s) = \sqrt{\frac{\sum_{i=1}^{N} (x_{i,f_s} -x_{i,500 Hz})^2}{N} }  ,
\end{equation}
\begin{equation}
NRMSD (f_s) = \frac{RMSD(f_s) }{x_{avg}},
\end{equation}
where $x_i$ is the $R_{reg}$ or $k$ at each SOC level and $f_s$ is the used sampling frequency. It can be observed that if the system is capable of sampling faster than 1 Hz, a 5 s interruption length is sufficient to obtain data with less than 5\% NRMSD. With a slower data acquisition system, for example 0.1 Hz, a result with around 10 \% NRMSD can be achieved by prolonging the interruption length to 60 s.

\begin{figure}
    \centering
    \includegraphics{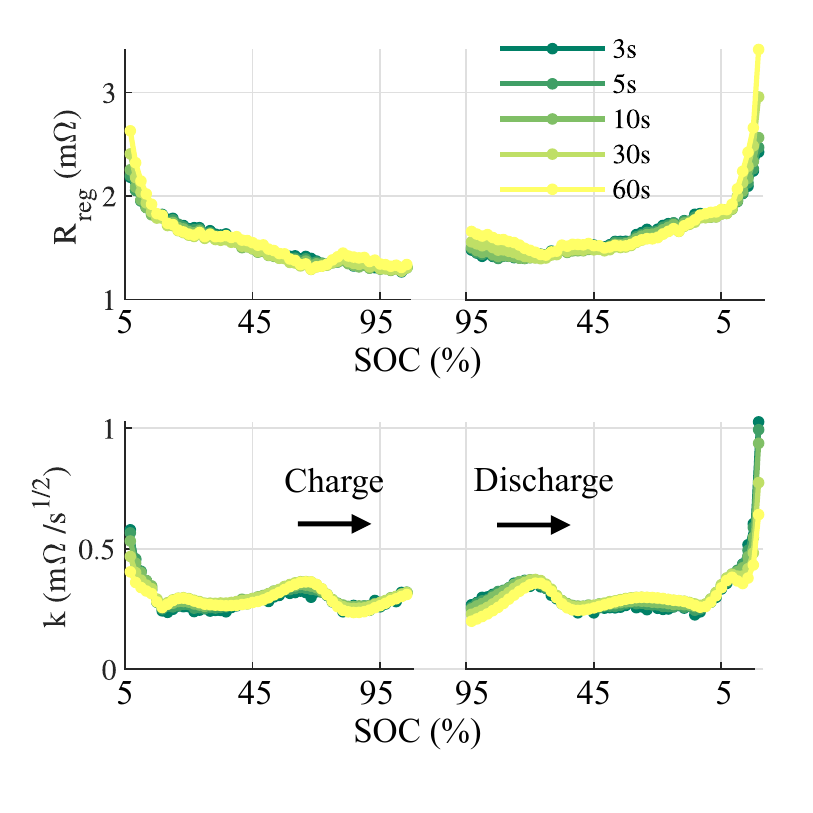}
    \caption{$R_{ref}$ and $k$ values obtained with different interruption lengths with 500 Hz sampling frequency.}
    \label{fig:Error_of_pulse_length_results}
\end{figure}

In Fig.~\ref{fig:RMSD_length_freq}, the results with 500 Hz in each interruption length case are used as the reference. However, the interruption length itself will introduce deviations in the results, as shown in Fig.~\ref{fig:Error_of_pulse_length_results}. It shows that a longer interruption length will lead to an overestimation in the $R_{reg}$ values and an underestimation in the $k$ values. This impact is more significant at the lower SOC ranges. 

\begin{figure}
    \centering
    \includegraphics{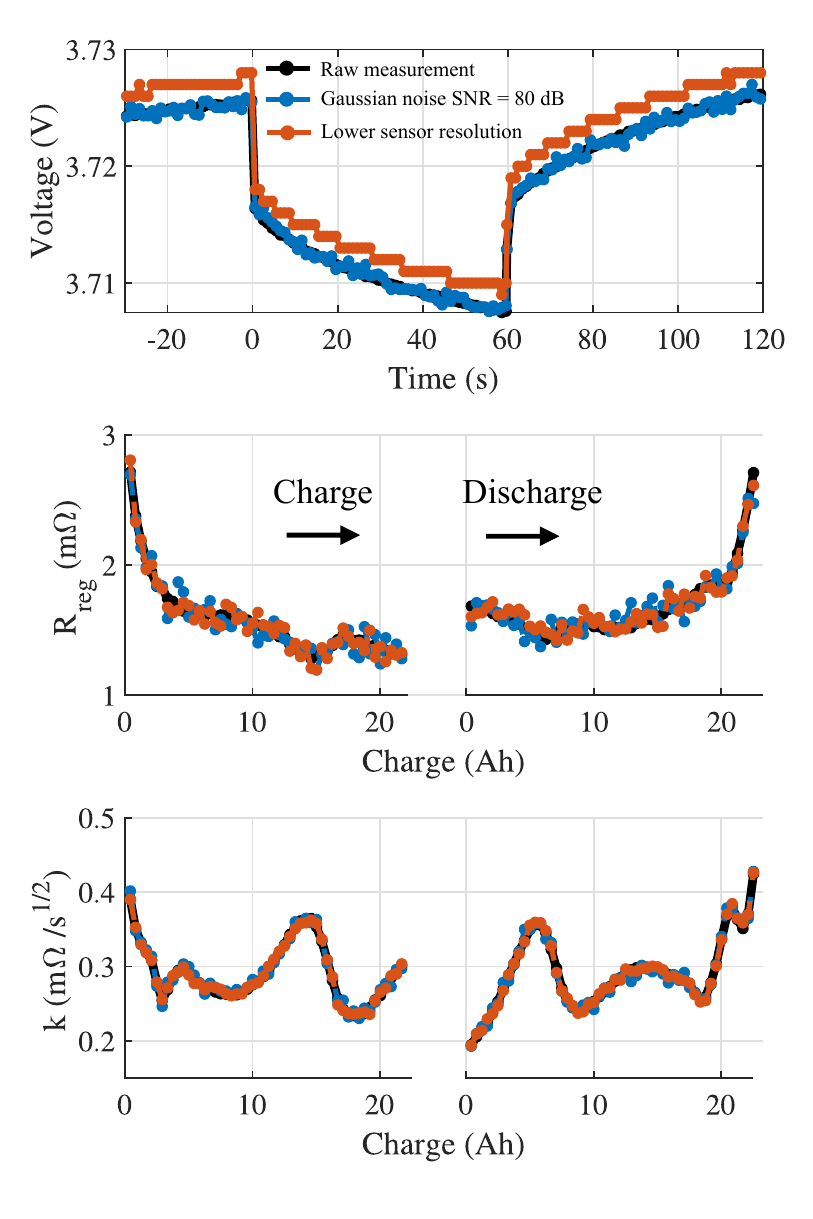}
    \caption{How the sensor resolution and environment noise affect the ICI method results.}
    \label{fig:Noise_sensor_result}
\end{figure}

Not only is the sampling frequency the limitation factor of the equipment in the applications, but also the sensor accuracy, resolution and the noise level in the environment. In the investigation, the measurement is down-sampled to 1 Hz first, and then in one case a Gaussian noise is added with 80 dB signal to noise ratio (SNR), while in the other case, the resolution of the voltage measured is reduced to 1 mV with a 2 mV offset. The results in Fig.~\ref{fig:Noise_sensor_result} show that the ICI method can be utilized with sensors with a lower resolution and can tolerate a certain level of noise.

This is a promising sign that the ICI method can be implemented with on-board sensors in battery systems for aging diagnostics, besides being used in laboratory tests. One example is to implement the ICI method in an electric vehicle during selected charging events. The hardware and software implementation are similar as for the pulse charging strategy, demonstrated in \cite{kalla2016generalized,bayati2020delivering}. A difference towards the pulse charging, which can be used with both fast charging and slow charging infrastructures, is that the ICI method is more suitable to use with slow charging to provide clear diagnostic information ($R_{reg}$, $k$, ICA and DVA curves). In principle, each time the vehicle is connected to a slow charger (most likely at home during the night), and the SOC is below a certain threshold, the user could chose to perform the ICI diagnostic during charging. Today BEVs have a battery pack ranging from 20 kWh to 100 kWh. On-board chargers utilising standard power outlets have a limited capacity, in Europe 3.7 kW (16 A singel-phase 230 V AC), in USA up to 6 kW \cite{sebastian2020adaptive}, meaning that a charging rate slower than C/5 is common. One factor favouring home charging is the price \cite{SmartAgent}, where the fast charging price leads to a driving cost similar to ICE vehicles whereas home charging to 1/3 or 1/4 of that. Furthermore, an indicator towards that slow charging will remain is the life time of the battery. In \cite{SmartAgent}, an increase of 78 \% of the battery resistance was reported after 120 fast charging events. Therefore, even in countries where home charging might be less common, e.g. China, it can definitely be favourable to use slow charging for overnight and over-day (work) parking. Regardless, the driver could always indicate when the car is being connected for charging, that the car is not needed for 6 hours, and the test can be automatically conducted. When the ICI method is implemented in the charging procedure, the charging time will be less than 3 \% longer (with the example of 10 s pulse every 5 mins during the constant current charge phase). Not only does it not disturb the drivers' usage, but also it can be beneficial for the battery lifetime, in a similar way as the pulse charging \cite{li2001effects,serhan2018effect}.

\section{Conclusions}

This work demonstrated the application of the intermittent current interruption method on commercial lithium ion batteries which has not been reported before. This method can extract most of the measurable electrical properties with a high repeatability. The parameters can be tracked during a battery life testing and have a very good correlation between the information extracted from the performed EIS measurements. It is noteworthy that the parameters $R_{reg}$ and $k$ from the ICI method are connected to physical processes and can therefore be used to characterise the battery degradation with different aging phenomena. The aging impact of the RPT sequence (including a capacity test, ICI test, 1 kHz AC impedance test, EIS test, charge and discharge pulse tests) is evaluated with a cell running the RPT continuously. One complete RPT can cause roughly 0.54 \% loss in the 1C discharge capacity. This impact can be reduced with a less frequent RPT schedule and a simplified RPT sequence, so the aging test can better target the specific case under investigation. 

Among the parameters that can be extracted from one ICI test, $R_{reg}$ and $k$ rely on a linear regression of the measurement data, and therefore these two parameters can be tracked easily even in a field test with limited sensor accuracy, signal to noise ratio and sampling frequency. A very interesting possibility introduced through the results in this article is that the ICI method has been shown to have the potential to be implemented in a charging sequence in electric vehicles and stationary storage applications, to track the battery aging properties.

\section*{Acknowledgment}
The authors would like to thank Energimyndigheten (P42789-1 and P45538-1) for the financing of this work.

\ifCLASSOPTIONcaptionsoff
  \newpage
\fi



%\bibliographystyle{IEEEtran}

\bibliographystyle{IEEEtran}
\bibliography{reference.bib}

\begin{IEEEbiography}[{\includegraphics[width=1in,height=1.25in,clip,keepaspectratio]{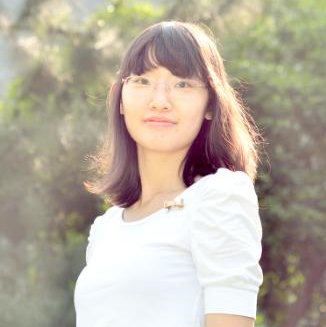}}]{Zeyang Geng}

Zeyang Geng is a PhD student at Chalmers University of Technology working with characterization, modelling and aging of lithium ion batteries. She took her B.Sc at Zhejiang University in 2013 and M.Sc at Chalmers University of Technology in 2015.
\end{IEEEbiography}

\begin{IEEEbiography}[{\includegraphics[width=1in,height=1.25in,clip,keepaspectratio]{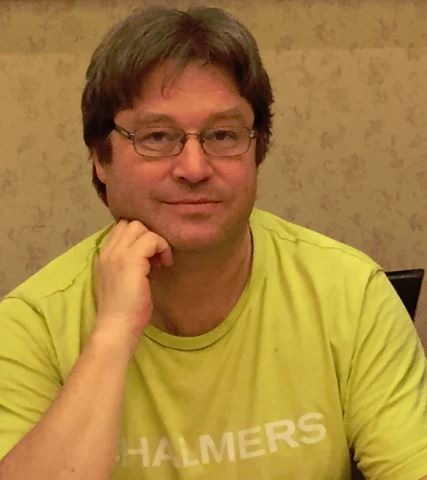}}]{Torbjörn Thiringer}

Torbjörn Thiringer works at Chalmers University of Technology, in Göteborg Sweden, as a professor in applied power electronics. He took his M.Sc and Ph.D at Chalmers University of Technology in 1989 and 1996 respectively. His areas of interest include the modeling, control and grid integration of wind energy converters into power grids, battery technology from cell modelling to system aspects, as well as power electronics and drives for other types of applications, such as electrified vehicles, buildings and industrial applications.
\end{IEEEbiography}

\begin{IEEEbiography}[{\includegraphics[width=1in,height=1.25in,clip,keepaspectratio]{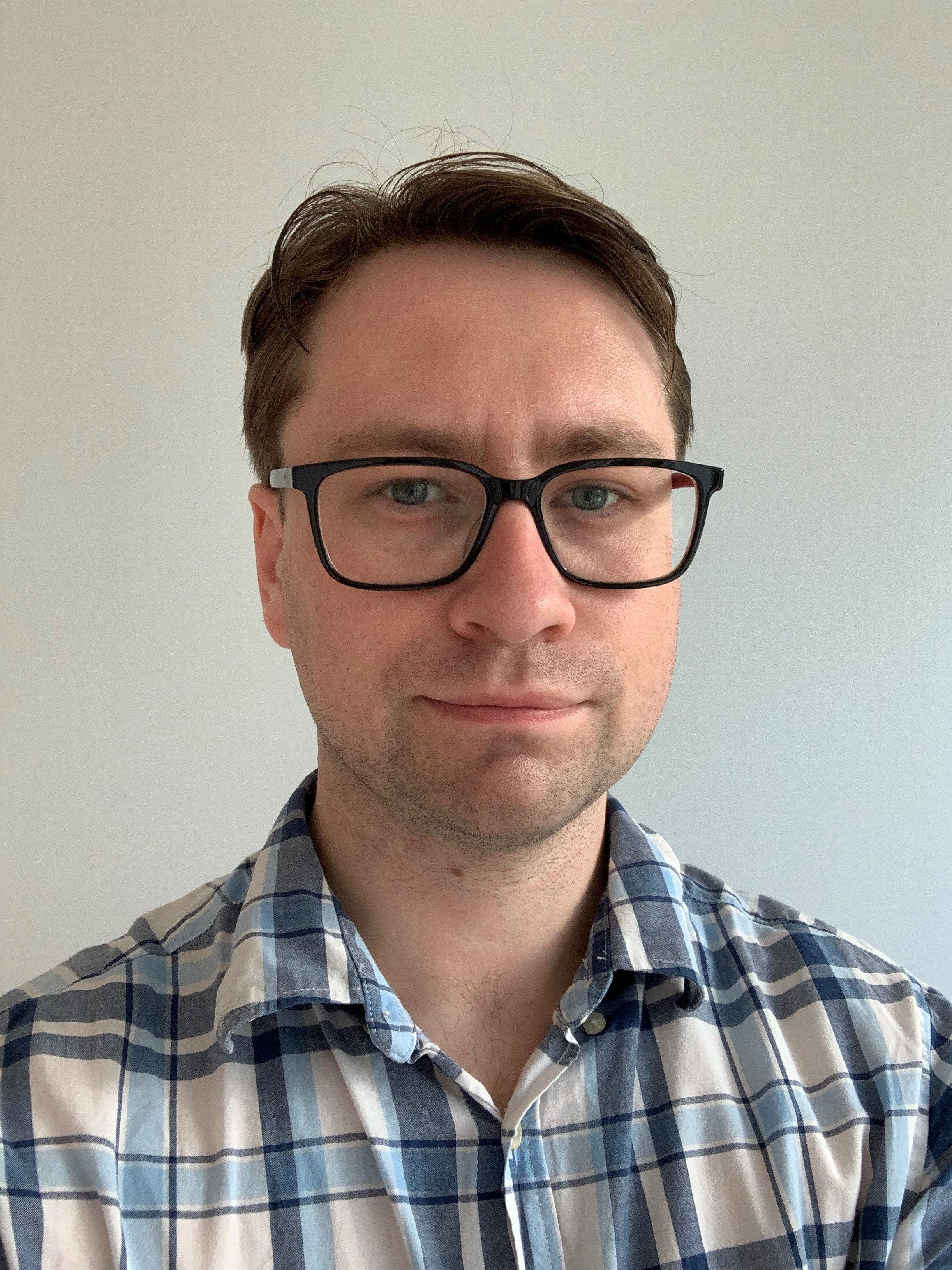}}]{Matthew J. Lacey}

Matthew Lacey received his MChem and PhD degrees from the University of Southampton in 2008 and 2012, respectively, with a focus on electrochemistry and new materials for lithium-ion batteries. He joined Uppsala University in 2012 to focus on lithium-sulfur battery chemistry, and in 2019 joined Scania CV AB as a Development Engineer in Materials Technology, with a focus on battery cell analytics, electrochemical methods and next-generation battery chemistries.
\end{IEEEbiography}

%\vfill


\end{document}